\documentclass[a4paper,11pt]{article}
\usepackage[toc,page]{appendix}
\usepackage{graphicx}
\usepackage{colortbl}
\usepackage{amsmath}
\usepackage{subfigure}
\usepackage{mathtools}
\usepackage{cite}

\DeclarePairedDelimiterX\braket[2]{\langle}{\rangle}{#1 \delimsize\vert #2}
\usepackage[utf8]{inputenc}
\usepackage{float}
\usepackage[normalem]{ulem}
\usepackage{epstopdf}
\usepackage[toc,page]{appendix}
\usepackage{amsfonts,amsmath,amssymb}
\usepackage{hyperref}

\usepackage{epsfig,multicol,bbm}
\usepackage{color}
\usepackage[dvipsnames]{xcolor}

\definecolor{darkblue}{rgb}{0.0, 0.0, 0.55}
\definecolor{grey}{rgb}{0.57, 0.64, 0.69}
\definecolor{lightbrown}{rgb}{0.71, 0.4, 0.11}

\newcommand{\be}{\begin{equation}}
\newcommand{\ee}{\end{equation}}

\date{}
 \newcommand{\rbm}[1]{{\color{red}\bf [Robb: #1]}}

\newcommand\fverb{\setbox\pippobox=\hbox\bgroup\verb}
\newcommand\fverbit{\egroup\item[\fbox{\unhbox\pippobox}]}

\newbox\pippobox

\begin{document}

\title{\bf Charged Soliton/Black Hole Phase Transitions in Higher-Order Lovelock Gravities}
\author{Seyed Naseh Sajadi$^{1}$\thanks{Electronic address: naseh.sajadi@gmail.com}\,,\,
Robert B. Mann$^{2}$\thanks{Electronic address: rbmann@uwaterloo.ca}\,,\,
Petarpa Boonserm$^{1}$\thanks{Electronic address: petarpa.boonserm@gmail.com}\,,\,\\
\small $^1$Department of Mathematics and Computer Science,
Faculty of Science, Chulalongkorn University, \\
\small Bangkok 10330, Thailand,\\
\small Department of Physics and Astronomy,
University of Waterloo, Waterloo, Ontario,\\\small N2L 3G1, Canada.\\
}

\maketitle

\begin{abstract}

We investigate the thermodynamics and phase structure of charged black holes in arbitrary-dimensional Lovelock--Maxwell theory. Using the covariant phase space formalism, we derive the conserved charges associated with spacetime symmetries and compute the thermodynamic quantities of static, spherically symmetric black hole solutions, including the temperature, Wald entropy, and the thermodynamic potentials conjugate to the Lovelock coupling constants and electric charge. We verify the first law of black hole thermodynamics and obtain the corresponding Smarr relation. We then study the phase transition between planar charged AdS black holes and AdS soliton solutions. The soliton geometry is constructed through a double analytic continuation of the planar black hole metric, accompanied by an analytic continuation of the electromagnetic charge. Working in the grand-canonical ensemble, we evaluate the Euclidean free energies of the competing configurations and analyze their dependence on the temperature, electric potential, and magnetic flux. Our results demonstrate the existence of black hole--soliton phase transitions and identify the regions of parameter space in which each phase is thermodynamically preferred. 

\end{abstract}

\section{Introduction}

Higher-order Lovelock gravity is a natural extension of general relativity to higher-dimensional spacetimes. It includes additional curvature terms while keeping many of the important properties of Einstein’s theory. Lovelock gravity has been shown to be  the most general metric theory of gravity in arbitrary dimensions that leads to second-order field equations \cite{Lovelock:1971yv,Lovelock:1972vz}. As a result, it avoids the ghosts and instabilities that usually appear in higher-derivative gravity theories. The Lovelock action is built from a finite number of curvature terms, each corresponding to a different power of the spacetime curvature. The lowest-order Lovelock terms are the cosmological constant and the Einstein-Hilbert term, the latter reproducing general relativity. The first higher-curvature correction is the Gauss-Bonnet term, and further corrections of cubic, quartic, and higher order appear naturally in dimensions greater than four. Each Lovelock term contributes to the field equations only above a specific spacetime dimension, making higher-order terms particularly relevant in higher dimensions \cite{Lovelock:1971yv,Lovelock:1972vz,Camanho:2015zqa,Zumino:1985dp,MullerHoissen:1985mm}.

There are several reasons why these theories are of great interest \cite{Padmanabhan:2013xyr,Deruelle:1989fj,Charmousis:2008kc}. In the context of string theory and other approaches to quantum gravity, higher-curvature terms arise as low-energy effective corrections to the Einstein–Hilbert action \cite{Zwiebach:1985uq,Boulware:1985wk}. Higher-order Lovelock terms significantly affect black hole properties, including horizon structure, and thermodynamics. These features make Lovelock gravity an important theoretical laboratory for testing ideas about holography, and the interplay between geometry and thermodynamics \cite{Cvetic:2001bk,Bravo-Gaete:2026hoe,Cai:2003gr, Brigante:2007nu,Crisostomo:2000bb,Aros:2000ij,Kofinas:2006hr, Ge:2008ni,Miskovic:2007mg,Miskovic:2008ck, Maeda:2011zn,Xu:2014tja,Hennigar:2015esa,Dykaarhun:2017mba}.
 In asymptotically anti-de Sitter (AdS) spacetimes, black holes exhibit a wide variety of thermodynamic behaviors and phase structures that have attracted considerable attention, especially in the context of the AdS/CFT correspondence \cite{Hawking:1982dh,Maldacena:1997re,Witten:1998zw,Witten:1998qj,Chamblin:1999hg,Chamblin:1999tk,Kubiznak:2012wp}.
  One of the best-known examples is the Hawking--Page phase transition between thermal AdS space and Schwarzschild--AdS black holes. In the AdS/CFT correspondence, this transition is interpreted as the confinement/deconfinement transition of the dual field theory \cite{Hawking:1982dh,Chamblin:1999tk,Kubiznak:2012wp,Kastor:2009wy,Frassino:2014pha}; it can also be interpreted as a solid/liquid phase transition
  \cite{Kubiznak:2014zwa}
in the context of black hole chemistry \cite{Kubiznak:2016qmn,Mann:2025xrb}.

  Beyond spherical horizons, planar and Ricci-flat AdS black holes also display a nontrivial phase transition between planar AdS black holes and AdS solitons \cite{Surya:2001vj}. By comparing their free energies it is possible to show that a transition from one to the other can take place depending on the parameters.  
 Related studies have also shown that electrically charged Ricci-flat black holes exhibit similar phase transitions, with the electric charge playing an important role in determining the thermodynamic phase structure \cite{Banerjee:2007by,Anabalon:2022ksf}. When gravity is coupled to non-linear electrodynamics, multiple phases and multicritical points become possible
\cite{Quijada:2023fkc}.
 
 The inclusion of higher-curvature interactions considerably enriches this picture. Recent studies have shown that Lovelock AdS black holes can also exhibit multicritical points and new thermodynamic phases \cite{Wu:2022plw,Suryaatmadja:2026dws}. 
 Motivated by these results, it is natural to study black hole-soliton phase transitions within the more general framework of Lovelock gravity. This helps us understand how electric charge, higher-curvature corrections, and the geometry of the horizon affect the thermodynamic behavior of gravitational systems. In particular, studying the effect of Lovelock coupling constants on the stability and preferred phase of black holes and solitons can provide deeper insight into holographic phase transitions and the impact of higher-curvature terms in quantum gravity.
 {One of our main findings is that all soliton-black hole phase transitions depend only on the parameters $(\alpha_0,\alpha_1)$, namely those characterizing Einstein--AdS gravity. Apart from their dependence on the spacetime dimension and on $L_{\mathrm{eff}}$, the thermodynamic properties of these solutions are universal across the entire Lovelock class.}

Our paper is organized as follows. In Section \ref{sec2}, we present the formalism of Lovelock theory in $D$ dimensions. We then derive the conserved charges of the theory using the covariant phase space method. We study the thermodynamics of black hole solutions in the theory and verify the first law of black hole thermodynamics as well as the Smarr formula. In section \ref{sec3}, we studied the soliton/black hole phase transition. Section \ref{seccon} is devoted to our conclusions.

%\rbm{Summary of results?}

\section{Thermodynamics of Black hole in Lovelock Gravity}\label{sec2}

The Lovelock action in $D$ dimensions is
\begin{equation}
I=\frac{1}{16\pi G_D}\int d^D x \,\sqrt{-g}\,
\left(\sum_{k=0}^{K} \alpha_k\, \mathcal{L}_k -4\pi G_{D}F_{ab}F^{ab}\right),
\end{equation}
where $K=\left\lfloor \frac{D-1}{2} \right\rfloor$ and $F=dA$ is the electromagnetic tensor. The Lovelock densities are given by
\begin{equation}
\mathcal{L}_k
=
\frac{1}{2^k}
\delta^{\mu_1\nu_1 \cdots \mu_k\nu_k}_{\rho_1\sigma_1 \cdots \rho_k\sigma_k}
\,
R^{\rho_1\sigma_1}{}_{\mu_1\nu_1}
\cdots
R^{\rho_k\sigma_k}{}_{\mu_k\nu_k},
\end{equation}
where $\delta^{\cdots}_{\cdots}$ is the generalized antisymmetric Kronecker delta:

\[
\delta^{\,c_{1} d_{1} \cdots c_{k} d_{k}}_{\,a_{1} b_{1} \cdots a_{k} b_{k}}
\;\equiv\;
(2k)!\;
\delta^{c_{1}}_{[a_{1}}
\delta^{d_{1}}_{b_{1}}
\cdots
\delta^{c_{k}}_{a_{k}}
\delta^{d_{k}}_{b_{k}]}.
\]

For Lovelock gravity coupled to a Maxwell field, the gravitational field equations can be written as
\begin{equation}
\mathcal{E}_{ab}
=
\mathcal{P}^{(k)}_{a}{}^{cde} R_{bcde}
-\frac{1}{2}g_{ab}\mathcal{L}_{(k)}
=
8\pi G\left(
F_{ac}F_{b}{}^{c}
-\frac{1}{4}g_{ab}F_{cd}F^{cd}
\right),
\end{equation}
where 
\begin{equation}
\mathcal{P}_{(k)}^{abcd}=
\frac{\partial \mathcal{L}_{(k)}}{\partial R_{abcd}}=
\frac{k}{2^{k}}
\delta^{ab a_2 b_2 \cdots a_k b_k}_{cd c_2 d_2 \cdots c_k d_k}
\prod_{i=2}^{k}
R_{a_i b_i}{}^{c_i d_i}
,\qquad \nabla_a \mathcal{P}_{(k)}^{abcd}=0.
\end{equation}
The terms involving covariant derivatives of $\mathcal{P}_{(k)}^{abcd}$ vanish identically, and the resulting gravitational field equations contain at most second derivatives of the metric.
The $D$-dimensional static solution is given by
\begin{align}\label{metric14}
ds^2 =& -f(r)\,dt^2 + \frac{dr^2}{f(r)} + r^2 \, d\Sigma_{D-2,\kappa}^2,\\
F=&\dfrac{\mathcal{Q}}{r^{D-2}}dt\wedge dr,
\end{align}
where $d\Sigma_{D-2,\kappa}^2$ denotes the metric in the unit $(D-2)$-dimensional space of constant curvature $\kappa$.
In Lovelock-Maxwell theory, the metric function $f(r)$ satisfies the algebraic equation

\begin{equation}\label{eqmetfunc}
\sum_{k=0}^{K}\hat{\alpha}_k
\left(\frac{\kappa-f(r)}{r^2}\right)^k
=
\frac{\mu}{r^{D-1}}-\dfrac{q^2}{r^{2D-4}} .
\end{equation}
Here  
\begin{equation}
    q^{2}=\dfrac{8\pi G}{(D-2)(D-3)}\mathcal{Q}^2.
\end{equation}
For large $r$, expanding the polynomial equation around the vacuum solution gives the first subleading correction \begin{equation}\label{eqmetfunc}
f(r)=\kappa+\frac{r^2}{L_{\mathrm{eff}}^2}
-\frac{M_{D}}{r^{D-3}}+\dfrac{Q_{D}^2}{r^{2D-6}}
+...,
\end{equation}
where
\begin{equation}\label{eqq15}
M_{D}=\frac{\mu}{\displaystyle
\sum_{k=1}^{K}k\,\hat{\alpha}_k
\left(-\frac{1}{L_{\mathrm{eff}}^2}\right)^{k-1}},\qquad Q^{2}_{D}=\frac{q^2}{\displaystyle
\sum_{k=1}^{K}k\,\hat{\alpha}_k
\left(-\frac{1}{L_{\mathrm{eff}}^2}\right)^{k-1}},\,\,\,\,\,\,\,\hat{\alpha}_k={\alpha}_k\,\dfrac{(D-3)!}{(D-2k-1)!}
\end{equation}
and where
\begin{equation}
    \sum_{k=0}^{K}\hat{\alpha}_{k}\left(-\frac{1}{L_{eff}^2}\right)^{k}=0
\end{equation}
defines
the effective length $L_{\mathrm{eff}}$.

For a diffeomorphism-invariant gravitational theory with a Lagrangian depending on the metric and the Riemann tensor, coupled to a Maxwell field, the symplectic potential current density can be written as
\begin{equation}
\Theta^{a}=\Theta^{a}_{\rm Lovelock}+\Theta^{a}_{\rm Maxwell}=2\sqrt{-g}\sum_{k=1}^{K}\alpha_{k}\mathcal{P}_{(k)}^{abcd}\nabla_{c}h_{bd}
-\sqrt{-g}F^{ab}\delta A_{b},
\end{equation}
where $h_{ab}=\delta g_{ab}$, $h=g^{ab}h_{ab}$. 
To evaluate the symplectic potential asymptotically, we use the leading-order expression for the Lovelock tensor $\mathcal{P}_{(k)}^{abcd}$:
\begin{equation}\label{eqPk19}
\mathcal{P}_{(k)}^{abcd}=\frac{k}{2}\frac{(D-2)!}{(D-2k)!}\left(-\frac{1}{L_{\rm eff}^{2}}\right)^{k-1}
\left(g^{ac}g^{bd}-g^{ad}g^{bc}\right)+\mathcal{O}\left(r^{-(D-1)}\right),
\end{equation}
where $L_{\rm eff}$ is the effective AdS curvature radius. Substituting this expression into the gravitational symplectic potential and retaining the leading asymptotic contribution gives
\begin{align}
\Theta^{a}_{\rm Lovelock}=\sqrt{-g}\sum_{k=1}^{K}\alpha_{k}k\frac{(D-2)!}{(D-2k)!}
\left(-\frac{1}{L_{\rm eff}^{2}}\right)^{k-1}\left(\nabla^{a}h-\nabla_{b}h^{ab}\right).
\end{align}

For the metric perturbations induced by variations of the mass parameter $M_D$ and the electric charge parameter $Q_D$, we consider
\begin{equation}
\begin{aligned}
h_{tt}=\frac{\delta M_D}{r^{D-3}}-\frac{2Q_D\delta Q_D}{r^{2D-6}},\qquad h_{rr}=
\frac{\delta M_D}{r^{D-3}f^{2}}-\frac{2Q_D\delta Q_D}{r^{2D-6}f^{2}},
\end{aligned}
\end{equation}
with the trace condition $h=g^{ab}h_{ab}=0.$ These perturbations can then be substituted into the asymptotic expression for $\Theta^{a}$ to determine the contribution of the gravitational sector to the conserved charges.
For the $r$ component, we have
\begin{align}
    \Theta^{r}_{\rm Lovelock} =&-\sum_{k=1}\sqrt{-g}\;\alpha_{k}{k}\,
\frac{(D-2)!}{(D-2k)!}\,
\left(-\frac{1}{L_{eff}^2}\right)^{k-1}\nabla_{r}h^{rr} \nonumber\\
=&-\sum_{k=1}\sqrt{\hat{\sigma}_{D-2,\,\kappa}}\;\alpha_{k}{k}\,
\frac{(D-2)!}{(D-2k)!}\,
\left(-\frac{1}{L_{eff}^2}\right)^{k-1}\delta M_{D},
\end{align}
\label{ThetaLove}
where
\[
\sqrt{-g}=r^{D-2}\,\sqrt{\hat{\sigma}_{D-2,\,\kappa}}.
\]
For the Maxwell term, we have
\begin{equation}
    \Theta^{r}_{\rm Maxwell}=-F^{rt}\delta A_{t}=-\sqrt{\hat{\sigma}_{D-2,\,\kappa}}\dfrac{\mathcal{Q}\delta \mathcal{Q}}{(D-3)r^{D-3}},\qquad \quad \delta A_{t}=\dfrac{\delta \mathcal{Q}}{(D-3)r^{D-3}}dt.
\end{equation}
Finally, using \eqref{eqq15} we obtain
\begin{align}
    \Theta^{r}=-
\sqrt{\hat{\sigma}_{D-2,\,\kappa}}\sum_{k=1}^{K}\alpha_{k}{k}\,\frac{(D-2)!}{(D-2k)!}\,
\left(-\frac{1}{L_{eff}^2}\right)^{k-1}{\delta M_{D}}
\end{align}
at large $r$, where we note that the symplectic potential of the Maxwell term does not contribute for $D>3$.

For Lovelock gravity, the Noether current associated with a vector field $\xi^a$ can be expressed as $J^a=\nabla_b Q^{ab},$ where the antisymmetric Noether charge using \eqref{eqPk19} is given by

\begin{equation}
    Q^{ab} = 2\sum_{k=1}^{K}\sqrt{-g}\,\alpha_{k}k\, \frac{(D-2)!}{(D-2k)!} \left(-\frac{1}{L_{\mathrm{eff}}^{2}}\right)^{k-1} \nabla^{[b}\xi^{a]} -{\frac14\,(\xi^c A_c)\,F^{ab}}.
\end{equation}
For the static spherically symmetric metric \eqref{metric14} with the timelike Killing vector $\xi^{a}=(\partial_{t})^{a}$, one can explicitly obtain
\begin{equation}
 Q^{a b}=
\sqrt{\hat{\sigma}_{D-2,\,\kappa}}\sum_{k=1}^{K}\alpha_{k}{k}\,
\frac{(D-3)!}{(D-2k)!}\,
\left(-\frac{1}{L_{eff}^2}\right)^{k-1} M_{D}.
\end{equation}
{The variation of the Hamiltonian
associated with $\xi^{a}$ is given by}
\begin{equation}
\delta H_\xi=\int_{\partial\Sigma}\Big(\delta Q^{ab}_\xi
-\xi^{[a}\Theta^{\,b]}(h)\Big)\, dS_{ab},
\end{equation}
explicitly
\begin{align}
\delta H_\xi^{\infty}=(D-2)\Sigma_{D-2}^{(\kappa)}\sum_{k=1}^{K}k\,\alpha_k
\frac{(D-3)!}{(D-2k-1)!}\left(-\frac{1}{L_{\mathrm{eff}}^2}\right)^{k-1}\;\delta M_D .
\end{align}
We identify the variation of the physical mass as $\delta \mathcal{M} = \delta H_\xi^{\infty}.$

 At the horizon, using $\nabla_a \xi_b = \mathcal{K} \,\epsilon_{ab}$, where \(\mathcal{K}\) is the surface gravity and \(\epsilon_{ab}\) is the binormal to the bifurcation surface, the charge variation receives contributions from both the Lovelock and electromagnetic sectors. It can be written as
\begin{align}
\delta H_\xi^{\mathcal{H}}={}-2 \int_{\mathcal{H}}\delta \mathcal{P}^{abcd}\,\nabla_c \xi_d\, dS_{ab}
-\int_{\mathcal{H}}\delta\left[(\xi^e A_e)F^{ab}\right]dS_{ab}
+\int_{\mathcal{H}}\xi^{[a}F^{b]c}\delta A_c\,dS_{ab}.
\end{align}
For a static, purely electric, spherically symmetric black hole, choosing the gauge $A_t(\infty)=0$, it becomes 
\begin{align}\label{eqphih}
\delta H_{\xi}^{\mathcal{H}}=-2\mathcal{K} \int_{\mathcal{H}}
\delta \mathcal{P}^{abcd}\,\epsilon_{cd}\,dS_{ab}+\Phi_H\,\delta \mathcal{Q},\qquad \Phi_H={\left(\dfrac{\delta \mathcal{M}}{\delta \mathcal{Q}}\right)_{S}=\dfrac{\Sigma_{D-2}^{(\kappa)}\mathcal{Q}}{(D-3)r_{h}^{D-3}}},
\end{align}
where $\Phi_H=A_t(\infty)-A_t(r_h)$ is the horizon electric potential.  Using the identity
\cite{Wald:1993nt}
\begin{equation}
S=-2\pi \int_{\mathcal{H}}\mathcal{P}^{abcd}\,\epsilon_{ab}\epsilon_{cd}\, dA,
\end{equation}
for the entropy,
we obtain
\begin{equation}
\delta H_\xi^{\mathcal{H}} = T\,\delta S+\Phi_H\,\delta \mathcal{Q},
\qquad
T = \frac{\mathcal{K}}{2\pi}.
\end{equation}
The Hamiltonian identity implies the following $\delta H_\xi^{\infty} = \delta H_\xi^{\mathcal{H}},$ thus
\begin{equation}
\delta\mathcal{M}=T\,\delta S+\Phi_H\,\delta \mathcal{Q} .
\end{equation}
Explicitly, the thermodynamic quantities are given as follows.
From \eqref{eqmetfunc}, one can get
\begin{equation}
\mu= \sum_{k=0}^{K}\dfrac{\hat{\alpha}_k \,\kappa^{k}}{r_h^{2k-D+1}} +\dfrac{q^2}{r_{h}^{D-3}},
\end{equation}
and the mass is given by
\begin{equation}\label{eqmass}
\mathcal{M} = \dfrac{(D-2)\Sigma_{D-2}^{(\kappa)}}{16\pi G}\left(\sum_{k=0}^{K}\dfrac{\hat{\alpha}_k \,\kappa^{k}}{r_h^{2k-D+1}} +\dfrac{q^2}{r_{h}^{D-3}}\right),
\end{equation}
the Hawking temperature is
\begin{equation}\label{eqtem}
T = \frac{f'(r_h)}{4\pi}=\frac{\displaystyle\sum_{k=0}^{K}(D-2k-1)\hat{\alpha}_k\kappa^k r_h^{-2k+1}-\dfrac{(D-3)q^2}{r_h^{2D-5}}}{\displaystyle4\pi\sum_{k=1}^{K}
k\hat{\alpha}_k\kappa^{k-1}r_h^{-2k+2}},
\end{equation}
and the entropy is \cite{Myers:1988ze}, \cite{Jacobson:1993xs} 
\begin{equation}\label{eqent}
    S=\dfrac{\Sigma_{D-2}^{(\kappa)}}{4G}\sum_{k=1}^{K}k\kappa^{k-1}{\hat{\alpha}_{k}\dfrac{(D-2)}{(D-2k)}}r_{h}^{D-2k}.
\end{equation}
By considering the coupling constants  as  thermodynamic quantities \cite{Kastor:2010gq},   the first law of thermodynamics is 
\begin{equation}
\delta \mathcal{M} = T\,\delta S+\Phi_H\,\delta \mathcal{Q}+\sum_{k=0}^{K}\phi_{k}\delta\hat{\alpha}_{k},
\end{equation}
where the conjugate potential  to the coupling constant $\hat{\alpha}_{k}$ is given by
%\rbm{Please check my correction below}\tcb{I agree. The entropy is also a function of $\hat{\alpha}$.}
%\begin{equation}
    %\phi_{k}= {\dfrac{(D-2)\Sigma_{D-2}^{(\kappa)}}{16\pi G}\kappa^{k-1} r_h^{D-2k}
    %\left[\frac{\kappa}{r_h} -\frac{4\pi k  T}{D-2k} \right]}
    %.
%\end{equation}

\begin{equation} 
\phi_k={\left(\frac{\delta\mathcal{M}}{\delta\hat{\alpha}_k}
\right)_{S,\mathcal{Q}}}= \frac{(D-2)\Sigma_{D-2}^{(\kappa)}}{16\pi G}\, \kappa^{k-1} r_h^{D-2k} \left[ \kappa-\frac{4\pi k\,T\,r_h}{D-2k} \right], \qquad k\geq 1. 
\end{equation}

Using dimensional scaling and applying Euler’s theorem, the Smarr formula is given by
\begin{equation}
{(D-3)\mathcal{M}=(D-2)TS+(D-3)\Phi_{H}\mathcal{Q}+\sum_{k=0}^{K}2(k-1)\phi_{k}\hat{\alpha}_{k}.}
\end{equation}

\section{Soliton/Black hole Phase Transitions}\label{sec3}
In the following, we study the planar soliton/planar black hole phase transition in the case $\alpha_{0}\neq 0$ and $q\neq 0$. Starting from the planar black hole $\kappa=0$ metric 
\begin{equation}
ds_{b}^2=-f_{b}(r)dt^2+\frac{dr^2}{f_{b}(r)}+r^2\left(d\psi^2+d\vec{x}_{D-3}^{\,2}\right),
\end{equation}
the Lovelock equation becomes
\begin{equation}\label{Loveeq}
\sum_{k=0}^{K}\hat{\alpha}_k\left(-\frac{f_{b}(r)}{r^2}\right)^k=\frac{\mu_{b}}{r^{D-1}}
-\frac{q_{b}^2}{r^{2D-4}}.
\end{equation}
At the event horizon $r=r_h$, we have $f_{b}(r_h)=0$ 
\begin{equation}
\mu_{b}=\alpha_{0}r_{h}^{D-1}+\dfrac{q_{b}^2}{r_{h}^{D-3}}.
\end{equation}

A candidate soliton geometry can be obtained by a double Wick
rotation of the planar black hole metric
\begin{equation}\label{Wickrotations}
t \rightarrow i\psi,\qquad \psi \rightarrow it,\qquad q\rightarrow iq_s.
\end{equation}
Choosing one of the planar coordinates, $\psi_s$, as the compact spatial direction,
the soliton metric takes the form 
\begin{equation}\label{eqsolitonmetric}
ds_s^2=-r^2dt^2+\frac{dr^2}{f_s(r)}+f_s(r)d\psi^2+r^2d\vec{x}_{D-3}^{\,2},
\end{equation}
here
\begin{equation}\label{Loveeq}
\sum_{k=0}^{K}\hat{\alpha}_k\left(-\frac{f_{s}(r)}{r^2}\right)^k=\frac{\mu_{s}}{r^{D-1}}+\frac{q_{s}^2}{r^{2D-4}},\qquad \mu_{s}=\alpha_{0}r_{s}^{D-1}-q_{s}^2/r_{s}^{D-3}.
\end{equation}
Regularity at $r=r_{s}$ requires the period of $\psi$ to be
\begin{equation}
\eta_s\equiv\Delta\psi=\frac{4\pi}{f_s'(r_s)}=\dfrac{4\pi\alpha_{1}}{\left[
\alpha_{0}(D-1)r_{s}+\frac{(D-3)q_s^2}{r_s^{2D-5}}
\right]}.
\label{eq:soliton_period}
\end{equation}

To compare the Euclidean actions of the black hole and the soliton,
we introduce a common radial cutoff $r=R$. The induced geometries
on the boundary at $r=R$ must be matched before subtracting the
corresponding Euclidean actions. Matching the proper lengths $(\beta_b,\beta_s)$ of the Euclidean time
directions gives
\begin{equation}
\beta_b\sqrt{f_b(R)}=\beta_s R,\qquad \sqrt{f_{s}}\eta_{s}=\eta_{b}R,
\label{eq:time_matching}
\end{equation}
where at the limit $R\to \infty$, one finds that 
\begin{equation}\label{eqmachcond}
    \beta_{b}=\beta_{s}=\beta,\qquad \eta_{b}=\eta_{s}=\eta .
\end{equation}
This matching procedure ensures that both geometries are evaluated under identical thermal and geometric boundary conditions. {To compute the regularized action, all higher-order Lovelock curvature terms contribute. However, here we consider only terms up to quadratic order.}
The Euclidean action for Einstein–Gauss–Bonnet (EGB) gravity with a Maxwell field is
\begin{equation}
I_E=I_E^{\rm bulk}+I_E^{\rm GHY}
+I_E^{\rm Myers}+I_E^{\rm ct},
\end{equation}
here the bulk term is given by
\begin{align}
I_E^{\rm bulk}=-\frac{1}{16\pi G_D}\int_{\mathcal M}d^Dx\,\sqrt{g}
\left[\alpha_{0}+\alpha_{1} R+\alpha_{2}\mathcal L_{2}
-4\pi G_D F_{\mu\nu}F^{\mu\nu}\right],
\end{align}
with
\begin{equation}
\mathcal L_{2}=R_{\mu\nu\rho\sigma}R^{\mu\nu\rho\sigma}
-4R_{\mu\nu}R^{\mu\nu}+R^2.
\end{equation}
The Gibbons--Hawking--York boundary term for Einstein gravity and the Myers boundary term for Gauss--Bonnet gravity are given by
\begin{align}
I_E^{\rm GHY}=&-\frac{\alpha_{1}}{8\pi G_D}
\int_{\partial\mathcal M}d^{D-1}x\,\sqrt{h}\,K,\\
I_E^{\rm Myers}=&-\frac{\alpha_{2}}{4\pi G_D}
\int_{\partial\mathcal M}d^{D-1}x\,\sqrt{h}
\left(J-2\widehat{G}_{ab}K^{ab}\right),
\end{align}
here,
\begin{equation}
J_{ab}=\frac{1}{3}\left(2KK_{ac}K^c{}_b
+K_{cd}K^{cd}K_{ab}-2K_{ac}K^{cd}K_{db}-K^2K_{ab}\right),\qquad J=h^{ab}J_{ab},
\end{equation}
where \(I_E^{\rm ct}\) denotes the counterterm contribution, if
needed, to regularize the action.
For the grand-canonical ensemble, the electric potential is fixed
at the boundary, and no additional Maxwell boundary term is required. Explicitly, the bulk term is given by
\[
\begin{aligned}
I_E^{\rm bulk}
=-\frac{\beta\eta\Sigma_{D-3,\kappa}}{16\pi G_D}
\int_{r_+}^{R}dr\,\bigg\{&\alpha_0 r^{D-2}+\alpha_1\bigg[
-r^{D-2}f''-2(D-2)r^{D-3}f'\\
&\qquad +(D-2)(D-3)r^{D-4}(\kappa-f)
\bigg]\\
&+2\alpha_2(D-2)(D-3)\bigg[(D-4)(D-5)r^{D-6}(\kappa-f)^2\\
&\qquad -2(D-4)r^{D-5}(\kappa-f)f' +r^{D-4}(f')^2-r^{D-4}(\kappa-f)f''
\bigg]-\nonumber\\
&4\pi G_D r^{D-2}F_{\mu\nu}F^{\mu\nu}\bigg\}\,dr.
\end{aligned}
\]

After integration over the boundary, the Myers term and  Gibbons--Hawking--York term become
\begin{equation}
\begin{aligned}
I_E^{\rm Myers}(R)=&-\frac{\beta\eta\Sigma_{D-3,\kappa}\alpha_{2}}
{4\pi G_D}\bigg[\frac{(D-2)(D-3)}{2}R^{D-4}f'(\kappa-f)
+\nonumber\\
&(D-2)(D-3)(D-4)R^{D-5}f\left(\kappa-\frac{f}{3}\right)
\bigg]_{r=R}.
\end{aligned}
\end{equation}

\[
I_{\mathrm{GHY}}=-\frac{\beta\eta\Sigma_{D-3,\kappa}\alpha_{1}}{8\pi G_D}
\left[(D-2)R^{D-3}f(R)+\frac{R^{D-2}}{2}f'(R)
\right].
\]

The counter-term of the Einstein part is given by

\[
I_{\mathrm{ct,Einstein}}=\frac{\alpha_1}{8\pi G_D}
\int_{\partial\mathcal M} d^{D-1}x\,\sqrt{h}\,
\frac{D-2}{L_{\mathrm{eff}}}.
\]

At a radial cutoff \(R\), the bulk action becomes
\begin{align}
I_{\mathrm{bulk}}^{(b)}(R)
={}&-\frac{\beta_b\eta_b\Sigma_{D-3}^{(0)}}{8\pi G_D}
\int_{r_h}^{R}dr\Bigg[-\frac{D-1}{L^2}r^{D-2}+(D-3)q_b^2r^{2-D}
\nonumber\\
&\quad
+2\alpha_2(D-2)(D-3)\frac{d}{dr}\left(r^{D-4}ff'+\frac{D-4}{2}r^{D-5}f^2\right)
\Bigg].
\end{align}
Performing the radial integration yields
\begin{equation}
\begin{aligned}
I_{\mathrm{bulk}}^{(b)}(R)
={}&\frac{\beta_b\eta_b\Sigma_{D-3}^{(0)}}{8\pi G_D}
\Bigg[\frac{R^{D-1}-r_h^{D-1}}{L^2}-q_b^2\left(r_h^{3-D}-R^{3-D}\right)
\\
&\quad
+2\alpha_2(D-2)(D-3)\left(R^{D-4}f(R)f'(R)+\frac{D-4}{2}R^{D-5}f(R)^2\right)
\Bigg].
\end{aligned}
\label{eq:bulk_black_hole}
\end{equation}
It should be noted that the Gauss--Bonnet bulk term vanishes at the horizon. Therefore, for $\kappa=0$, the thermodynamics of the black hole depends only on $\alpha_0$ and $\alpha_1$. The Myers term cancels the Gauss--Bonnet bulk term. Therefore, the Gibbons--Hawking-York boundary term is
\begin{equation}
I_{\mathrm{GHY}}^{(b)}=-\frac{1}{8\pi G_D}\int_{\partial\mathcal M}d^{D-1}x\,\sqrt{h}\,K=-\frac{\beta_b\eta_{b}\Sigma_{D-3}^{(0)}}{8\pi G_D}
\left[\frac{(D-1)R^{D-1}}{L^2}-\frac{(D-1)\mu_b}{2}+\frac{q_b^2}{R^{D-3}}\right],
\label{eq:EM_GHY}
\end{equation}
and the leading AdS counterterm for a flat boundary is
\begin{equation}
I_{\mathrm{ct}}^{(b)}
=\frac{1}{8\pi G_D}
\int_{\partial\mathcal M}d^{D-1}x\,\sqrt{h}\,
\frac{D-2}{L}=\frac{\beta_b\eta_{b}\Sigma_{D-3}^{(0)}}{8\pi G_D}
\Bigg[\frac{(D-2)R^{D-1}}{L^2}-\frac{(D-2)\mu_b}{2}\\
+\frac{(D-2)q_b^2}{2R^{D-3}}+\cdots\Bigg].
\label{eq:EM_counterterm}
\end{equation}
Similarly, for the soliton solution geometry, the Gibbons--Hawking-York term and counterterm are given by
\begin{equation}
\begin{aligned}
I_{\mathrm{GHY}}^{(s)}(R)=-\frac{\beta_s\eta_s\Sigma_{D-3}^{(0)}}{8\pi G_D}
\left[\frac{(D-1)R^{D-1}}{L^2}-\frac{(D-1)\mu_s}{2}
-\frac{q_s^2}{R^{D-3}}\right],
\end{aligned}
\label{eq:GHY_soliton}
\end{equation}
\begin{equation}
\begin{aligned}
I_{\mathrm{ct}}^{(s)}(R)=\frac{\beta_s\eta_s\Sigma_{D-3}^{(0)}}{8\pi G_D}
\Bigg[\frac{(D-2)R^{D-1}}{L^2}-\frac{(D-2)\mu_s}{2}-\frac{(D-2)q_s^2}{2R^{D-3}}+O(R^{-D+1})\Bigg].
\label{eq:ct_soliton_expansion}
\end{aligned}
\end{equation}
The regularized Euclidean actions are given by
\begin{equation}
I_{\mathrm{reg}}^{(b)}=-\frac{\beta_b\eta_{b}\Sigma_{D-3}^{(0)}}{16\pi G_D}
\left[\frac{r_h^{D-1}}{L^2}+\frac{q_b^2}{r_h^{D-3}}\right],\qquad
I_{\mathrm{reg}}^{(s)}=-\frac{\beta_s\eta_s\Sigma_{D-3}^{(0)}}{16\pi G_D}
\left[\frac{r_s^{D-1}}{L^2}-\frac{q_s^2}{r_s^{D-3}}\right].
\label{eq:regularized_action_soliton}
\end{equation}
In this paper, we study the thermodynamics of the system in the grand canonical ensemble.
A state in the grand canonical ensemble is characterized by the inverse temperature \(\beta\) and the electric potential \(\Phi\). In this ensemble, the respective Gibbs free energies are  given by
\begin{align}
    G_{b}=&\dfrac{I_{\mathrm{reg}}^{(b)}}{\beta_{b}}=-\frac{\eta_{b}\Sigma_{D-3}^{(0)}}{16\pi G_D}
\left[\frac{r_h^{D-1}}{L^2}+\frac{q_b^2}{r_h^{D-3}}\right]=-\frac{\eta_{b}\Sigma_{D-3}^{(0)}}{16\pi G_D}
\left[\frac{r_h^{D-1}}{L^2}+\dfrac{8\pi G_{D}(D-3)r_{h}^{D-3}\Phi_{b}^{2}}{(D-2){\left(\Sigma_{D-2}^{(0)}\right)^2}}\right],\label{freeenergyaction}\\
G_{s}=&\dfrac{I_{\mathrm{reg}}^{(s)}}{\beta_{s}}=-\frac{\eta_{s}\Sigma_{D-3}^{(0)}}{16\pi G_D}
\left[\frac{r_s^{D-1}}{L^2}-\frac{q_s^2}{r_s^{D-3}}\right]=-\frac{\eta_{s}\Sigma_{D-3}^{(0)}}{16\pi G_D}\left[\frac{r_s^{D-1}}{L^2}-\dfrac{8\pi G(D-3)r_{s}^{D-3}\Phi_{s}^{2}}{(D-2)\eta_{s}^{2}{\left(\Sigma_{D-2}^{(0)}\right)^2}}\right].\label{freeenergyaction1}
\end{align}
Using the above relations, we can compute the variables of the system as follows
\begin{align}
T_{b}=&\frac{(D-1)\hat{\alpha}_0 r_{h}
-\dfrac{(D-3)q_{b}^2}{r_h^{2D-5}}}
{4\pi\hat{\alpha}_1},\quad
S_{b}=\frac{\Sigma_{D-2}^{(0)}\alpha_1}{4G}
r_h^{D-2},\quad
\mathcal{M}_{b}=
\frac{(D-2)\Sigma_{D-2}^{(0)}}{16\pi G}
\left(\hat{\alpha}_0 r_h^{D-1}+\frac{q_{b}^2}{r_h^{D-3}}
\right),\nonumber\\  \Phi_b=&\frac{{\Sigma_{D-2}^{(0)}}\mathcal{Q}_{b}}{(D-3)r_h^{D-3}},\qquad
\Phi_s=\frac{{\Sigma_{D-2}^{(0)}}\mathcal{Q}_s\eta_s}{(D-3)r_s^{D-3}}.
\end{align}
The difference between the free energies after using the matching conditions \eqref{eqmachcond} is
\begin{align}
    \Delta G=&-\frac{\Sigma_{D-3}^{(0)}\eta}{16\pi G_D}
\Bigg[\frac{r_h^{D-1}}{L^2}
+\frac{q_b^2}{r_h^{D-3}}-\frac{r_s^{D-1}}{L^2}-\frac{q_s^2}{r_s^{D-3}}
\Bigg]\nonumber\\
=&-\dfrac{\Sigma_{D-3}^{(0)}\eta}{16\pi G_{D}}\Bigg[\dfrac{r_{h}^{D-1}}{L^2}-\dfrac{r_{s}^{D-1}}{L^2}+\dfrac{8\pi G_{D}(D-3)}{(D-2){\left(\Sigma_{D-2}^{(0)}\right)^2}}\left(r_{h}^{D-3}\Phi_{b}^2+\dfrac{r_{s}^{D-3}\Phi_{s}^{2}}{\eta^{2}}\right)\Bigg],
\end{align}
here 
\begin{equation}
    \dfrac{1}{L^2}=\dfrac{\alpha_{0}}{\alpha_{1}(D-1)(D-2)}.
\end{equation}
For an asymptotically AdS spacetime with $L^2>0$, we require $\alpha_0>0$ and $\alpha_1>0$ in the Einstein limit. The black hole/soliton phase transition occurs when $\Delta {G}=0$. From the regularity condition \eqref{eq:soliton_period}, we have
\begin{equation}
    r_{s}=\dfrac{4\pi\alpha_{1}\pm\sqrt{\left(4\pi\alpha_{1}\right)^2-\dfrac{32\pi\alpha_{0}(D-1)(D-3)^2\Phi_{s}^2}{(D-2){\left(\Sigma_{D-2}^{(0)}\right)^2}}}}{2(D-1)\alpha_{0}\eta_{s}}=\dfrac{2\pi\alpha_{1}}{(D-1)\alpha_{0}\eta}\left[1\pm\sqrt{1-\dfrac{\Phi_{s}^2}{\Phi_{0}^{2}}}\right],
\end{equation}
here $\Phi_{0}^2=\pi\alpha_{1}^2(D-2){\left(\Sigma_{D-2}^{(0)}\right)^2}/(2\alpha_{0}(D-1)(D-3)^2)$. For $\alpha_{0}>0$ and $\alpha_{1}>0$ (asymptotic AdS), the $(+)$ branch gives positive $r_{s}>0$. From the temperature $T=1/\beta$ of the black hole, we have
\begin{align}
r_h=&\frac{4\pi\hat{\alpha}_1 T+\sqrt{(4\pi\hat{\alpha}_1 T)^2+
\dfrac{32\pi (D-1)(D-3)^2\hat{\alpha}_0\Phi_b^2}{(D-2){\left(\Sigma_{D-2}^{(0)}\right)^2}}}}{2(D-1)\hat{\alpha}_0}\nonumber\\
=&\dfrac{2\pi\alpha_{1}}{(D-1)\alpha_{0}\beta}\left[1+\sqrt{1+\dfrac{2(D-1)(D-3)^2\alpha_{0}\beta^2\Phi_{b}^{2}}{\pi\alpha_{1}^2(D-2){\left(\Sigma_{D-2}^{(0)}\right)^2}}}\right].
\end{align}

In order to work in the grand-canonical ensemble, we must express the free energy in terms of the fixed boundary conditions, namely, the common temperature, the value of \(\eta\), the chemical potential, and the magnetic flux.
Therefore, we introduce the following dimensionless quantities:
\begin{equation}
    X=\dfrac{\Phi_{s}}{L\Sigma_{D-2}^{(0)}},\qquad Y=\dfrac{\eta \Phi_{b}}{L\Sigma_{D-2}^{(0)}},\qquad\delta=\dfrac{\beta}{\eta},\qquad\lambda=\dfrac{L}{\eta},\qquad x_{h}=\dfrac{r_{h}}{\eta},\qquad x_{s}=\dfrac{r_{s}}{\eta}
\end{equation}
in turn yielding
\begin{align}
    x_{h}=&\dfrac{2\pi\lambda^2}{(D-1)^2(D-2)\delta}\left[1+\sqrt{1+\dfrac{2(D-1)^2(D-3)^2\delta^2Y^2}{\pi\alpha_{1}}}\right]=\dfrac{a\lambda^2 \mathcal{A}}{\delta},\\
    x_{s}=&\dfrac{2\pi\lambda^2}{(D-1)^2(D-2)}\left[1+\sqrt{1-\dfrac{2(D-1)^2(D-3)^2 X^2}{\pi\alpha_{1}}}\right]=a\lambda^2 \mathcal{B},
\end{align}
and the free energy is given by
\begin{align}
    \delta G=-\dfrac{\Sigma_{D-3}^{(0)}\eta^{D}}{16\pi L^2}\left[x_{h}^{D-1}-x_{s}^{D-1}+\dfrac{8\pi (D-3)\lambda^{4}}{D-2}\left(x_{h}^{D-3}Y^{2}+x_{s}^{D-3}X^{2}\right)\right].
\end{align}
In dimensionless form, we have
\begin{align}
    \delta \hat{G}=\dfrac{16\pi \delta G}{\Sigma_{D-3}^{(0)}\eta^{D-2}\lambda^{2D-4}a^{D-1}}=\mathcal{B}^{D-1}-\dfrac{\mathcal{A}^{D-1}}{\delta^{D-1}}-\dfrac{8\pi(D-3)}{(D-2)a^{2}}\left(\dfrac{\mathcal{A}^{D-3}Y^{2}}{\delta^{D-3}}+\mathcal{B}^{D-3}X^{2}\right).
\end{align}

In what follows, we analyze the signs of \(\delta G\) as a function of the three independent rescaled thermodynamic variables \(X\), \(Y\), and \(\delta\). In the special case $X=Y=0$, we obtain
\begin{equation}
    \delta\hat{G}=1-\frac{1}{\delta^{D-1}},
\end{equation}
which corresponds to the case of an uncharged black hole and an uncharged soliton. In this case, setting $\delta=1$ yields $\delta\hat{G}=0$, indicating that the two solutions have equal Gibbs free energies. For $Y=0$, the transition temperature
occurs when
\begin{equation} 
\delta^{-1} =\dfrac{\mathcal{B}}{2} \left[ 1- \frac{8\pi(D-3)X^2}{(D-2)a^2 \mathcal{B}^2} \right]^{\frac1{D-1}}.
\end{equation}

\begin{figure}
\centering
\includegraphics[width=0.3\columnwidth]{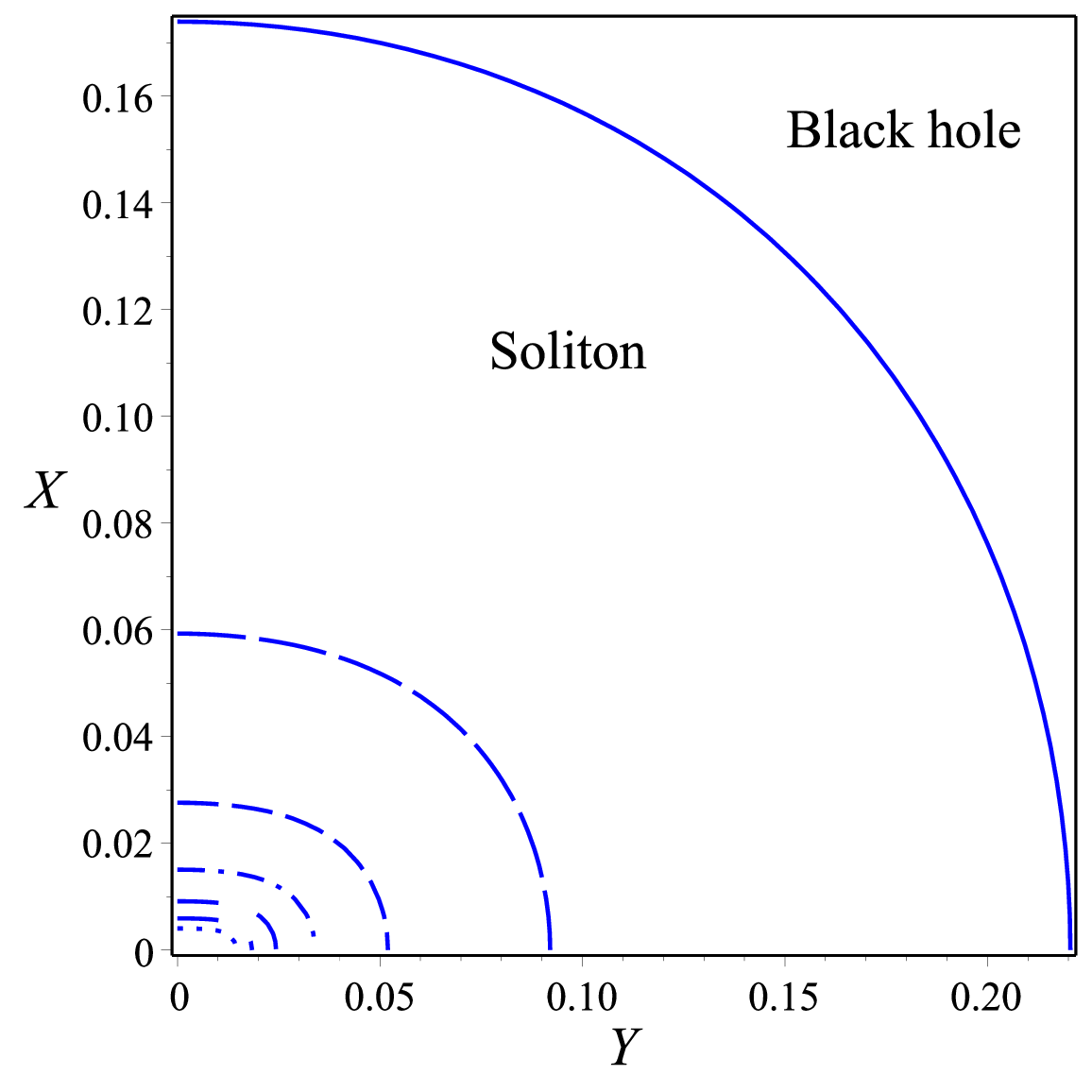}
%\includesvg[width=0.5\columnwidth]{plotgyw.svg}
\includegraphics[width=0.3\columnwidth]{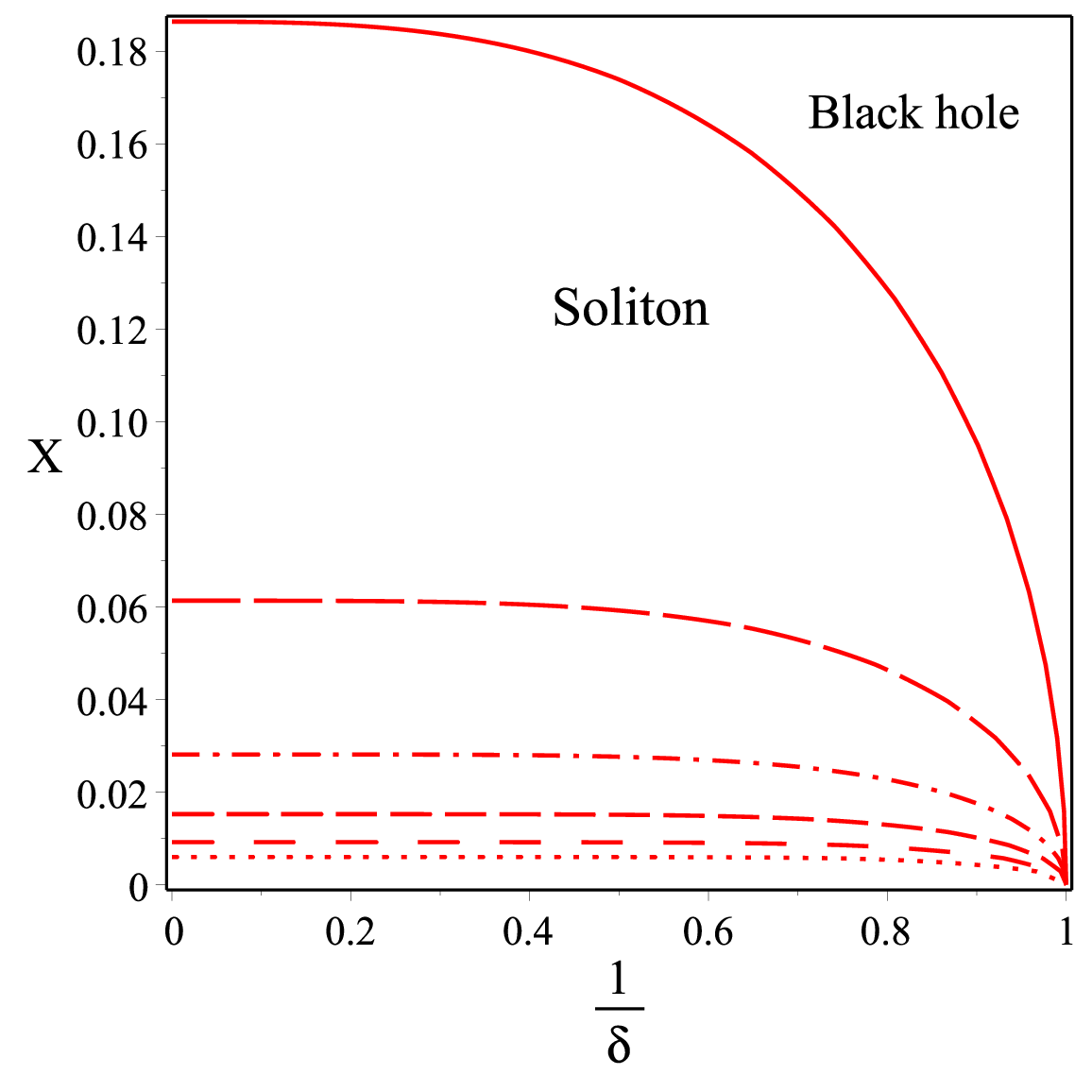}
\includegraphics[width=0.3\columnwidth]{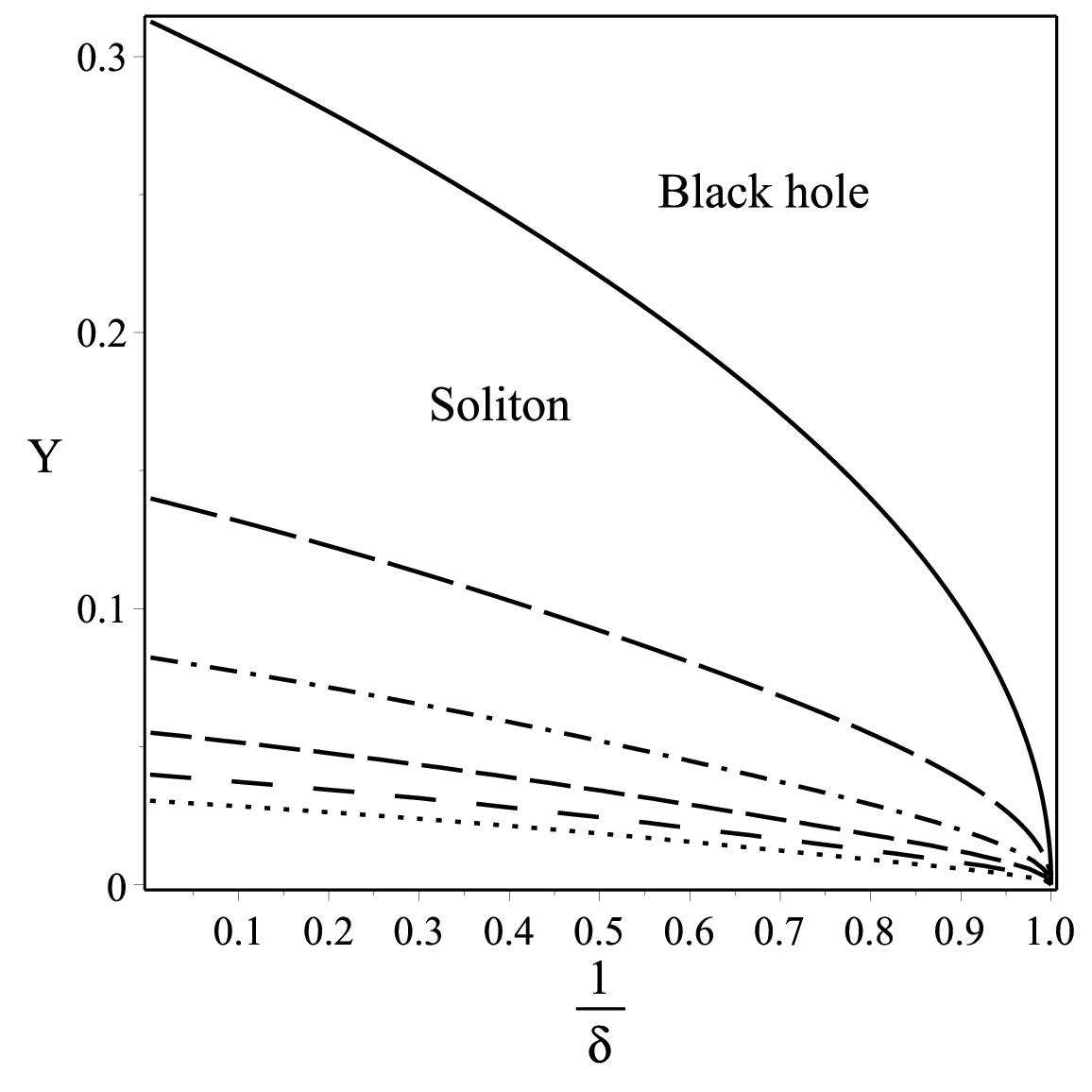}
\caption{Phase transition curves between charged planar AdS black holes and charged planar solitons. Along the curves, the free energies of the two solutions are equal, and the two phases coexist. Below the curves, the soliton phase is thermodynamically preferred, whereas above the curves, the black-hole phase dominates. The curves correspond to $D=4,5,6,7,8,9$, ordered from top to bottom. Left panel: $X$ as a function of $Y$ for $\delta^{-1}=0$. Middle panel: $X$ as a function of $\delta^{-1}$ for $Y=0$. Right panel: $Y$ as a function of $\delta^{-1}$ for $X=0$.}
\label{fig:phase_transition}
\end{figure}

We first consider $D=4$ \cite{Banerjee:2007by,Anabalon:2022ksf}. 
In Fig.~\ref{fig:phase_transition}, the phase transition curves illustrate the thermodynamic competition between charged planar AdS black holes and charged planar solitons as the boundary thermodynamic parameters vary. Along each curve, the Gibbs free energies of the two solutions are equal, indicating that they are thermodynamically degenerate. These curves separate the regions in which the black hole and soliton phases are thermodynamically preferred. Below the curves, the Gibbs free energy of the soliton is lower, and hence the soliton phase dominates. Conversely, above the curves, the black hole phase has lower Gibbs free energy and is therefore thermodynamically preferred. Along the curves, the two phases coexist.

The curves in Fig.~\ref{fig:phase_transition} correspond to spacetime dimensions $D=4,5,6,7,8,9$, ordered from top to bottom. This ordering demonstrates the dependence of the phase boundaries on the spacetime dimension and highlights the role of dimensionality in the thermodynamic competition between the black hole and soliton phases.

In the case of $\kappa=0$, in the absence of the cosmological constant and the electric charge, $\hat{\alpha}_0=0$ and $q=0$, the equation governing the static black hole geometry becomes the following:
\begin{equation}
\sum_{k=1}^{K}\hat{\alpha}_k
\left(-\frac{f(r)}{r^2}\right)^k
=\frac{\mu}{r^{D-1}}.
\label{eq:planar_lovelock}
\end{equation}
For a black hole with a regular horizon at $r=r_h>0$, the horizon condition is $f(r_h)=0$. Substituting this condition into Eq.~\eqref{eq:planar_lovelock} yields $\mu=0$. Therefore, under these assumptions, the field equation does not admit a planar black hole solution with a nonzero mass parameter. Under the simultaneous conditions $\kappa=0$, $\hat{\alpha}_0=0$, and $q=0$, the field equation does not provide the required nonzero-mass black hole solution. Consequently, the phase-transition analysis cannot be applied directly to this case.

{Using the thermodynamic quantities \eqref{eqphih}, \eqref{eqmass}, \eqref{eqtem}, and \eqref{eqent} for a planar black hole (\(\kappa=0\)), we find that the Gibbs free energy 
\begin{equation}
G_b=\mathcal{M}-TS-\Phi\,\mathcal{Q}, 
\label{Gbeqn}
\end{equation} 
exactly reproduces the result obtained from the Euclidean action in \eqref{freeenergyaction}. Since all contributions from the higher-order Lovelock couplings vanish in the planar case, this agreement holds for arbitrary Lovelock order \(K\geq 2\). Consequently, if the free energy is defined thermodynamically through the above relation, then, by means of the double analytic continuation, the corresponding soliton free energy is given by \eqref{freeenergyaction1} for any \(K\). Therefore, the phase structure derived in this section is universal for Lovelock theories coupled to Maxwell electrodynamics, independently of the Lovelock order. In particular, the phase transition in Fig.~\ref{fig:phase_transition} remains unchanged for arbitrary \(K\).}

\section{Conclusion}\label{seccon}

We have investigated the thermodynamic properties and phase structure of black hole solutions in arbitrary-dimensional Lovelock--Maxwell theory. Using the covariant phase space formalism, we derived the conserved charges associated with the spacetime symmetries and obtained the complete set of thermodynamic quantities, including the temperature, entropy, and the thermodynamic potentials conjugate to the electric charge and Lovelock coupling constants. Our analysis confirmed the validity of the first law of black hole thermodynamics and yielded the corresponding Smarr relation. 

Motivated by previous studies of AdS black hole--soliton transitions, we further explored the phase structure of planar solutions. The charged soliton geometry was constructed through a double analytic continuation of the planar black hole solution, allowing us to compare the thermodynamic properties of the two competing configurations. Employing the Euclidean action approach and evaluating the associated free energies, we demonstrated the existence of phase transitions between the charged planar black hole and charged soliton phases {for $K=2$, namely in Einstein-Gauss-Bonnet gravity.} Working in the grand-canonical ensemble, we analyzed how the phase structure depends on the temperature, electric potential, and magnetic flux. We found that the transition temperature is sensitive to both the electric and magnetic sectors, leading to a rich phase diagram in which either the black hole or the soliton can become thermodynamically dominant. In the limit of vanishing magnetic flux, our results reproduce previously known charged black hole--soliton transitions and recover the critical behavior reported in earlier studies. 

{Finally, our results extend to arbitrary Lovelock order $K$. Using the Gibbs free energy \eqref{Gbeqn}, which reproduces the Euclidean action result for the $K=2$ Einstein--Gauss--Bonnet theory, we find that all contributions from higher-order Lovelock couplings vanish. Consequently, the soliton--black hole phase structure depends only on the Einstein--AdS parameters $(\alpha_0,\alpha_1)$ and is therefore universal across all Lovelock theories coupled to Maxwell electrodynamics. Apart from their dependence on the spacetime dimension and the effective AdS radius $L_{\rm eff}$, the thermodynamic properties of these solutions are  identical to those of Einstein--AdS gravity. A natural question for further research is whether similar results can be found in more general higher-curvature theories of gravity.}

%\tcr{Finally, our results extend to all orders in $K$.  This is easily seen by employing the expression \eqref{Gbeqn} for the Gibbs free energy, which agrees with the Euclidean action calculation for the $K=2$ Einstein-Gauss-Bonnet theory.
%We find for any value of $K$ that all contributions from the higher-order Lovelock couplings vanish. Hence,
 %the phase structure derived in this paper is universal for Lovelock theories  coupled to Maxwell electrodynamics, independently of the Lovelock order $K$.}
%Our findings extend the analysis of black hole--soliton phase transitions beyond Einstein gravity to the broader framework of Lovelock theories, highlighting the role played by higher dimensions in gravitational thermodynamics. 

 %Finally, we note that 
 %all soliton/black hole phase transitions depend only on the parameters $(\alpha_0,\alpha_1)$, namely the parameters of Einstein-AdS gravity. Apart from a dependence on the number of dimensions and on $L_{\rm eff}$, the thermodynamic behavior of these objects is the same for all Lovelock theories.  {It would be interesting to see to what extent this situation holds for other higher-curvature theories of gravity.}

%\tcr{\bf Such a transition is absent in conformal gravity, which can be checked explicitly from this paper:2609.15737}

\section*{Acknowledgements} 

This research project is supported by the Second Century Fund (C2F), Chulalongkorn University and by the Natural Sciences and Engineering Research Council of Canada.

\end{document}